\documentclass[structabstract]{aa}
\usepackage{graphicx}
\usepackage{txfonts}
\usepackage{subfigure}
\usepackage{multirow}
\usepackage{natbib}
\usepackage{appendix}
\usepackage{array}
\usepackage{soul}
\begin{document}
   \title{High-Resolution X-ray Spectroscopy of the Interstellar Medium}

   \subtitle{\textit{XMM-Newton} Observation of the LMXB \object{GS 1826$-$238}}

   \author{C. Pinto
          \inst{1}
          \and
          J. S. Kaastra\inst{1,2}
          \and
          E. Costantini\inst{1}
          \and
          F. Verbunt\inst{2}
          }

   \institute{SRON Netherlands Institute for Space Research,
              Sorbonnelaan 2, 3584 CA Utrecht, The Netherlands\\
              \email{c.pinto@sron.nl}
         \and
             Astronomical Institute, Utrecht University,
             P.O. Box 80000, 3508 TA Utrecht, The Netherlands\\
             }

   \date{Received April 21, 2010 ; accepted July 13, 2010}

 
  \abstract
   {} 
   {The interstellar medium (ISM) has a multiphase structure characterized by gas, dust and molecules. The gas can be found in different charge states: neutral, low-ionized (warm) and high-ionized (hot). It is possible to probe the multiphase ISM through the observation of its absorption lines and edges in the X-ray spectra of background sources.}
   {We present a high-quality RGS spectrum of the low-mass X-ray binary GS 1826$-$238 with an unprecedent detailed treatment of the absorption features due to the dust and both the neutral and ionized gas of the ISM. We constrain the {column density} ratios within the different phases of the ISM and measure the abundances of elements such as O, Ne, Fe and Mg.}
   {We found significant deviations from the proto-Solar abundances: oxygen is over-abundant by a factor $1.23 \pm 0.05$, neon $1.75 \pm 0.11$, iron $1.37 \pm 0.17$ and magnesium $2.45 \pm 0.35$. The abundances are consistent with the measured metallicity gradient in our Galaxy: {the ISM appears to be metal-rich in the inner regions}. The spectrum also shows the presence of warm/hot ionized gas. The gas column has a total ionization degree less than 10\%. We also show that dust plays an important role {as expected from the position of GS 1826$-$238}: most iron appears to be bound in dust grains, while 10$-$40\% of oxygen consists of a mixture of dust and molecules.}
   {}


   \maketitle
%

\section{Introduction}
\label{sec:introduction}

The interstellar medium of our Galaxy (ISM) is a mixture of dust and gas in the form of atoms, molecules, ions and electrons. It manifests itself primarily through obscuration, reddening and polarization of starlight and the formation of absorption lines in stellar spectra, and secondly through various emission mechanisms (broadband continuum and line emission). The gas is found in both neutral and ionized phases \citep[for a review, see][]{ferriere}. The neutral phase is a blend of {\sl cold molecular gas} ($T \sim 20-50$ K), found in the so called dark clouds, and {\sl cold atomic gas} ($T \sim 100$ K) inherent in the diffuse clouds, while the {\sl warm atomic gas} has temperatures up to $10^4$ K. The atomic gas is well traced by \ion{H}{i} and mainly concentrated in the Galactic plane with clouds up to few hundreds pc above it. The {\sl warm ionized gas} is a low-ionized gas, with a temperature of $\sim 10^4$ K. It is mainly traced by H$\alpha$ line emission and pulsar dispersion measures; it can reach a vertical height of 1 kpc. The {\sl hot ionized gas} is characterized by temperatures of about $10^{6}$ K. It is heated by supernovae and stellar winds from massive stars; it gives rise to high-ionization absorption lines and the soft X-ray background emission. The study of the ISM is very interesting because of its connection with the evolution of the entire Galaxy: stellar evolution enriches the interstellar medium with heavy elements, while the ISM acts as source of matter for the star forming regions.

High resolution X-ray spectroscopy has become a powerful diagnostic tool for constraining the chemical and physical properties of the ISM. Through the study of the X-ray absorption lines in the spectra of background sources it is possible to probe the various phases of the ISM of the Galaxy. First of all, the K-shell transitions of low-Z elements, such as oxygen and neon, and the L-shell transitions of iron fall inside the soft X-ray energy band. Secondly, the presence of different charge states for each element allows us to constrain the multiphase ISM, e.g. its ionization state and temperature distribution.

\citet{Schattenburg} first measured ISM absorption edges in the X-ray band using the \textit{Einstein Observatory} and they found features consistent with the \ion{O}{i} $1s-2p$ line and traces of \ion{O}{ii}. After the launch of the {\textit{XMM-Newton}} and \textit{Chandra} satellites a new era for the study of the ISM opened up. The grating spectrometers on board of these satellites, RGS and LETGS/HETGS respectively, provide a spectral resolution high enough to resolve the main absorption edges and lines. Recently, \cite{yao09} found high-ionization absorption lines of ions such as \ion{O}{vi} to \ion{O}{viii} and \ion{Ne}{viii} to \ion{Ne}{x} in the HETGS spectrum of the low-mass X-ray binary \object{Cyg X-2}, and argued that the bulk of the \ion{O}{vi} should originate from the conductive interface between the cool and the hot gas. Other work has revealed complex structure around the oxygen K-shell absorption edge \citep{paerels, devries, JuettI}. \citet{costantini2005} argued that the feature of the scattering halo of Cyg X-2 near the \ion{O}{i} K-edge can be attributed to the presence of dust towards the source, with a major contribution from silicates such as olivine and pyroxene. In their paper on \object{Sco X-1}, observed with \textit{XMM-Newton}, \citet{costantini} found clear indications of extended X-ray absorption fine structures (EXAFS) near the absorption edge of oxygen.

In this work we report the detection of absorption lines and edges in the high-quality spectrum of the low-mass X-ray binary (LMXB) \object{GS 1826$-$238} obtained by the \textit{XMM-Newton} Reflection Grating Spectrometer \citep[RGS, ][]{denherder01}. In order to constrain the continuum parameters we also used the {EPIC-pn} \citep{struder01} dataset of this source. \citet{Thompson08} using \textit{XMM-Newton} and RXTE observations on April 2003 derived a high unabsorbed bolometric flux $F \sim 3.5 \times 10^{-12} \,  {\rm W \, m}^{-2}$. The source is well suited for the analysis of the ISM also because of its column density $N_{\rm H} \sim 4 \times 10^{25} \,  {\rm m}^{-2}$ (see Table \ref{table:rgs_fit}), high enough to produce prominent O and Fe edges. We assume the distance of the source to be $6.1 \pm 0.2$ kpc \citep{heger}.

We analyze the absorption in the spectrum as follows. We first remove the bursts, because they add a strongly variable component to the spectrum. Then we determine the source continuum by fitting simultaneously EPIC and RGS data. In second instance we use only the high-resolution RGS spectra to constrain the absorption contributions. We search for statistically significant features by adding several absorbers in sequence: cold gas, warm gas, hot gas, dust and molecules. All of these appear to be important.

\begin{table}
\caption{Observations used in this paper. We report the total exposure length together with the net exposure time remaining after screening of the background and removal of bursts.}             
\label{table:1}      
\centering                          
\begin{tabular}{c c c c c}        
\hline\hline                 
ID & Date & Length & RGS & PN\\    
   &      &  (ks)  &  (ks)   &  (ks)  \\
\hline                        
   0150390101 & 2003 April 6 & 108 & 67.8 & 63.8 \\      
   0150390301 & 2003 April 8 & 92  & 77.8 & 67.8 \\
\hline                                   
\end{tabular}
\end{table}


\section{Observations and data reduction}
\label{data}

GS 1826$-$238 (Galactic coordinates $l=9^{\circ}.27$, $b=-6^{\circ}.09$) has been observed twice with \textit{XMM-Newton} for a total length of 200 ks (see Table \ref{table:1} for details). The data are reduced with the \textit{XMM-Newton} Science Analysis System (SAS) version 9.0.1.

GS 1826$-$238 is a bursting LMXB with a regular time separation between the bursts. Because the primary aim of the \textit{XMM-Newton} observations was the study of the bursts, the EPIC-pn detector was operated in timing mode, which means that imaging is made only in one dimension, along the RAWX axis. Along the row direction (RAWY axis), data from a predefined area on one CCD chip are collapsed into a one-dimensional row for a fast read-out. Then source photons are extracted between RAWX values $30-45$ and background photons are extracted between rows $2-16$, as recommended by the standard procedure.

\begin{figure}
   \centering
   \includegraphics[bb=70 370 550 710, width=8.5cm]{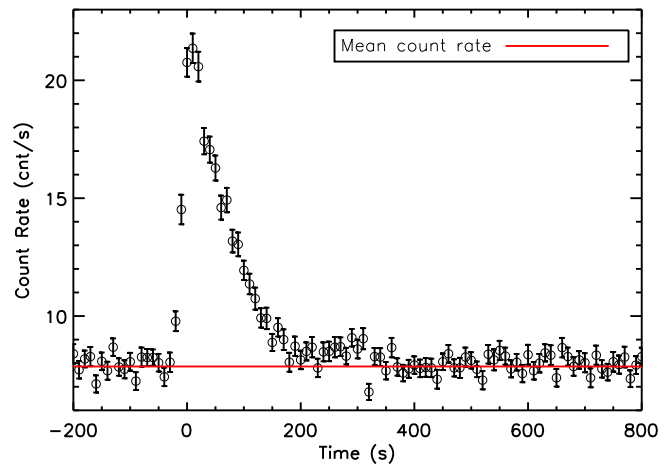}
      \caption{Mean profile of the bursts in the RGS lightcurve of the first observation. The zero point of the time scale is centered on the burst profile peak. The red line represents the mean count rate of the 2 ks around the peaks.}
         \label{fig:bursts}
   \end{figure}

We produce pn lightcurves mainly to remove the bursts intervals and to extract the spectra of the persistent part of the lightcurve. In the first observation 9 bursts have been detected, in the second observation 7 bursts. We plot the burst profiles of these 16 bursts in Fig. \ref{fig:bursts}. We estimate a mean duration of about 300 s for the bursts, and we remove for each burst 50 s before the peak to 250 s after it. Recently \cite{zand} suggested a mean duration of about 1 ks for the bursts, but they also argued that after the first 100 s the inferred emission decreases sharply by at least an order of magnitude, contributing only about 3\% to the fluence in the burst. After 250 s the flux of the burst has decreased by almost 2 orders of magnitude and its profile merges with the persistent lightcurve. {Thus, by removing 300 s for each burst, we retain less than $\sim$ 1\% burst emission, which is negligible compared to the persistent emission}.

We process RGS data with the SAS task \textit{rgsproc}. We produce the lightcurves for the background in CCD9 following the XMM-SAS guide\footnote{http://heasarc.nasa.gov/docs/xmm/abc/} in order to remove soft proton flares and spurious events. We create good time intervals (GTI) by removing intervals with count rates higher than 0.5 c s$^{-1}$. We reprocess the data again with \textit{rgsproc} by filtering them with the GTI for background screening and bursts removal. We extract response matrices and spectra for the two observations. The final net exposure times are reported in Table \ref{table:1}.

Our analysis focuses on the $7-31$ {\AA} ($0.4-1.77$ keV) first order spectra of the RGS detector. In order to fit the spectral continuum properly, we also use the $0.5-10$ keV EPIC spectra of both observations. We perform spectral analysis with SPEX\footnote{www.sron.nl/spex} version 2.01.05 \citep{kaastraspex}. We scale elemental abundances to the proto-Solar abundances of \cite{Lodders}{: N/H = 7.943$\times 10^{-5}$, O/H = 5.754$\times 10^{-4}$, Ne/H = 8.912$\times 10^{-5}$, Mg/H = 4.169$\times 10^{-5}$, Fe/H = 3.467$\times 10^{-5}$}. We use the C-statistic throughout the paper and adopt $1 \sigma$ errors.

   \begin{figure}
      \subfigure{ 
      \includegraphics[bb=145 220 460 570, width=4.3cm]{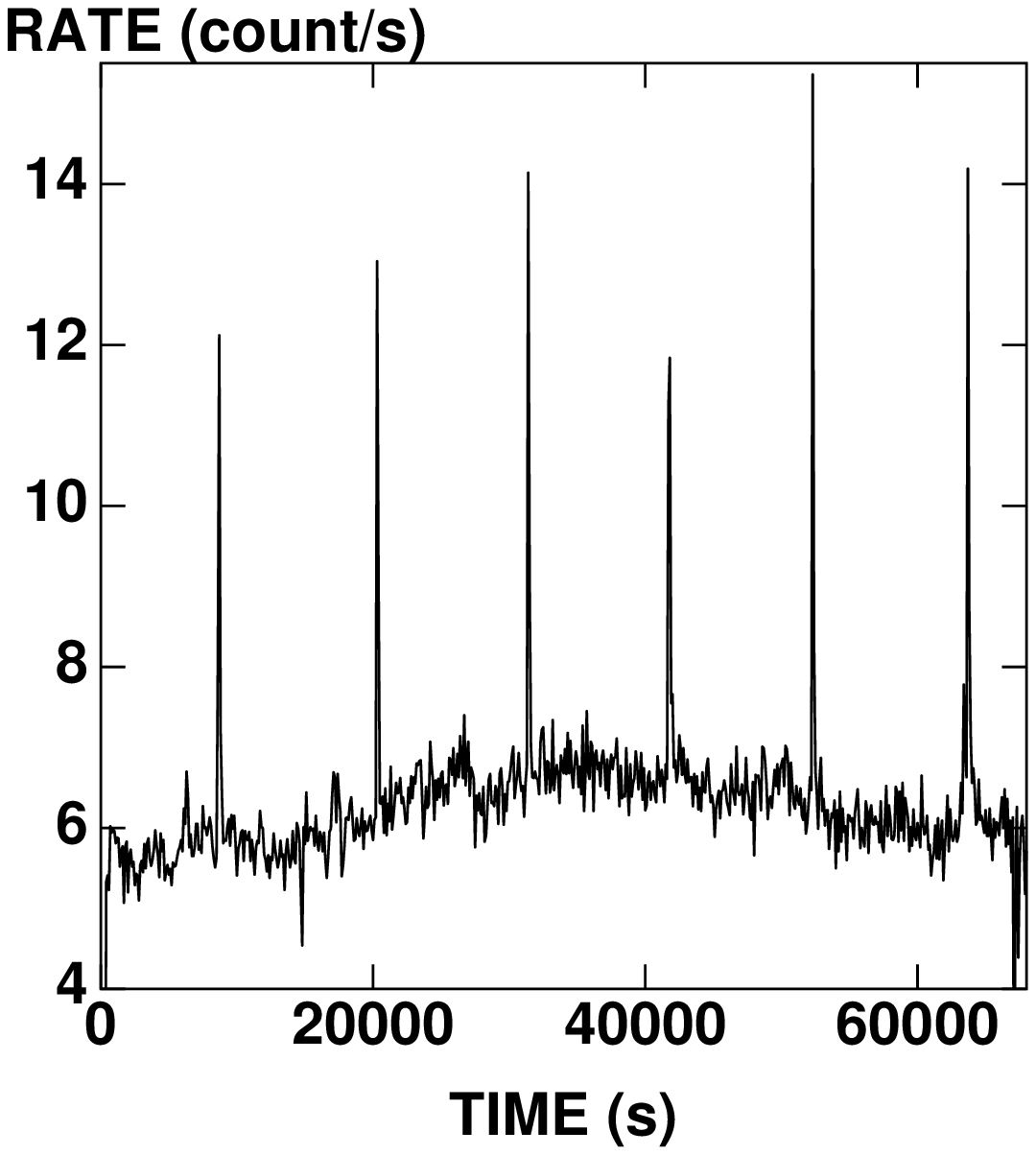}}
      \subfigure{ 
      \includegraphics[bb=145 220 460 570, width=4.3cm]{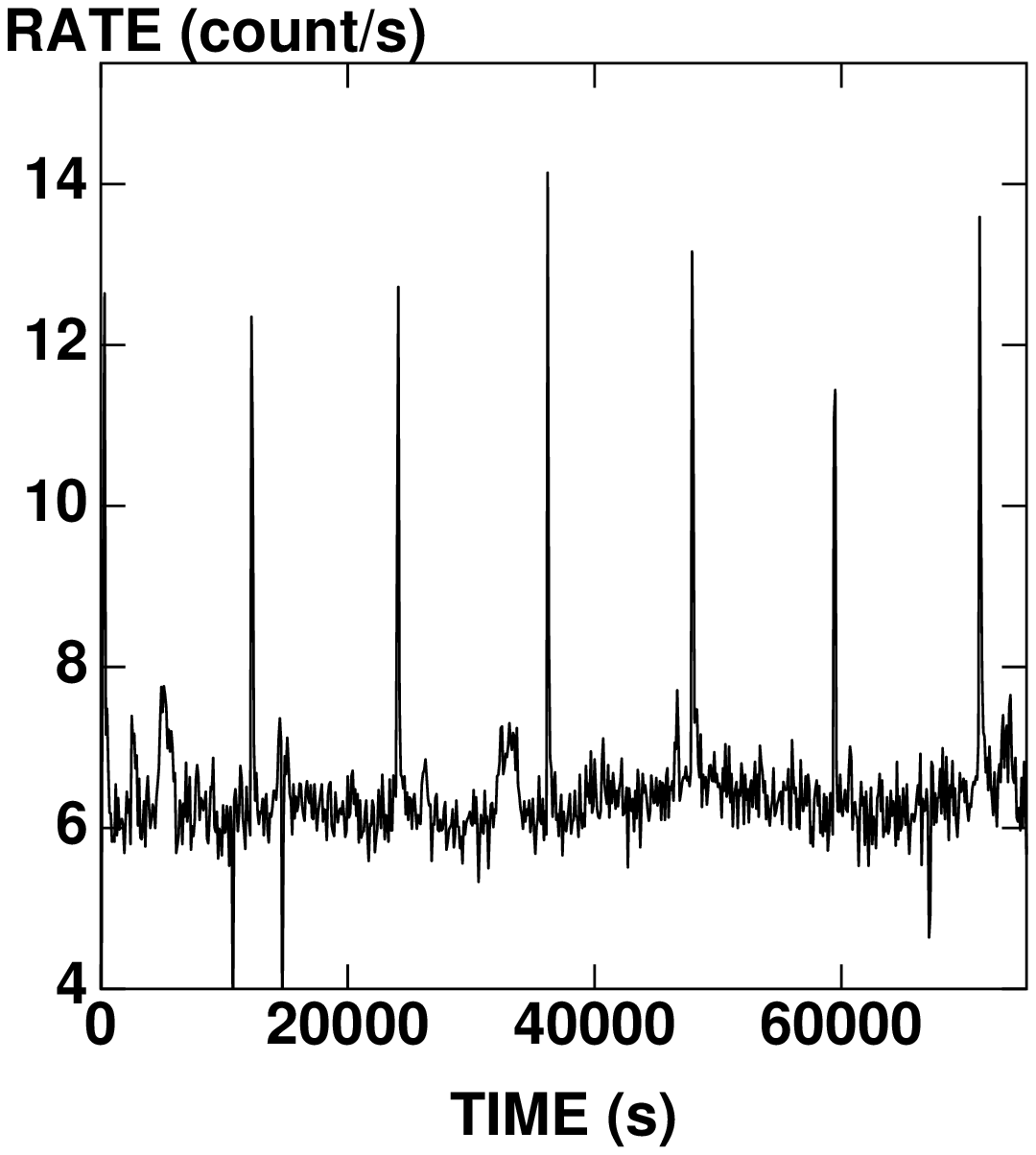}}
      \caption{{Lightcurves of the first (left) and the second (right) observation with RGS1. Intervals with high background have already been taken out.The bursts are still shown for displaying purpose, but their contribution in the spectra are removed as described in the text. The plot shows the quasi-periodicity of the bursts and illustrates why GS 1826$-$238 is called the 'clock-burster' LMXB \citep{ubertini99}.}}
          \label{fig:rgs_lcs_obs1}
    \end{figure}

\section{Spectral Modeling}
\label{analysis}

The first step of the spectral analysis consists of the determination of the continuum emission and the dominant absorption component. The best way to do this is to fit the spectra of RGS and EPIC-pn simultaneously. The \textit{XMM-Newton} cross-calibration is very complex, not only because of the different energy bands, but mainly because of their different features: RGS is sensitive in the soft X-ray energies with high spectral resolution, showing narrow absorption features, while pn has a low spectral resolution, therefore blurring the absorption features seen with RGS. EPIC-pn has a higher count rate compared to RGS and a broader energy band. The original RGS spectra are binned by a factor of 10 in this simultaneous fit. This is necessary to temporarily remove the narrow features due to the absorption lines. The pn spectra are resampled in bins of about 1/3 of the {spectral resolution (FWHM $\sim$ 50$-$150 eV between 0.5$-$10 keV)}, which is the optimal binning for most spectra.

A better local fit for absorption edges and lines is obtained from a separate RGS fit. In the RGS local fit we rebin the spectra only by a factor of two, {i.e. about 1/3 FWHM (the first order RGS spectra provide a resolution of 0.06$-$0.07 {\AA}}). This gives at least 10 counts/bin and a bin size of about 0.02 {\AA}.

\subsection{Simultaneous EPIC$-$RGS fits}
\label{sec:sim_fits}

\begin{figure}
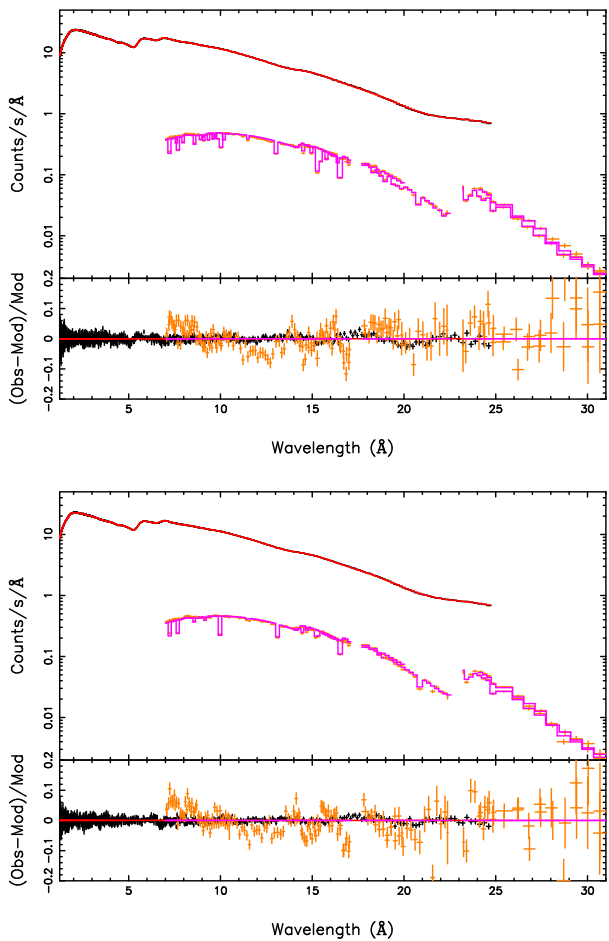

\begin{center}
      \subfigure{ 
      \includegraphics[width=6cm, angle=-90]{AAA_XMM_HBCC_obs1_lodd_fullepic_ang.ps}}
      \vspace{0.cm}
      \subfigure{ 
      \includegraphics[width=6cm, angle=-90]{AAA_XMM_HBCC_obs2_lodd_fullepic_ang.ps}}
\end{center}
      \caption{Simultaneous RGS$-$EPIC best fits of the first (top) and the second (bottom) observations. The model used is Model A (see Sect. \ref{sec:sim_fits} and Table \ref{table:fit}). The upper panels show from top to bottom the EPIC and RGS spectrum, respectively. The lower panels show the fit residuals (dark points with small error bars: EPIC). The dips present in the count spectrum of the RGS correspond to bad columns in the RGS with lower sensitivity.}
          \label{fig:epic_rgs_fit}
    \end{figure}

At first we follow the spectral modelling of \cite{Thompson08}. The continuum spectrum is modeled by emission from a black body and two comptonization models. The black body component arises from the thermal emission of the accretion disk around the neutron star. The first comptonization component (hereafter C1) describes the energy gain of the disk soft photons by scattering in the accretion disk corona. The second comptonization component C2 corresponds to scattered seed photons originating from regions closer to the NS surface, i.e. the boundary layer. \cite{Thompson08} applied a neutral absorber to the continuum mentioned above and fitted \textit{XMM-Newton}, Chandra and RXTE data. For this purpose we use the {\sl absm} model in SPEX: the model calculates the continuum transmission of neutral gas with cosmic abundances as published by \cite{Morrison}. In our case the same model does not give a satisfactory fit, especially around the neon and oxygen edges, and cannot fit the \ion{O}{i} line. This could be expected because the \cite{Morrison} model does not take into account the absorption lines and the possible variations in the abundances. Therefore we replace the {\sl absm} component with a {\sl hot} component, which describes the transmission through a layer of collisionally ionized plasma. At low temperatures it calculates the absorption of (almost) neutral gas (for further informations see the SPEX manual). We leave the temperature and {the O, Ne, Mg and Fe abundances of this absorber }free in the fit. {In the fits we ignore two small regions ($17.2-17.7$ {\AA} and $22.7-23.2$ {\AA}), close to the iron and oxygen edges respectively. The presence of dust and molecules affects the fine structure of the edge, thus these regions will be analyzed with more complex models in Sect. }\ref{sec:dust_analysis}{. However the ISM abundances are determined by the depth of the absorption edges, thus ignoring these small regions we still can constrain the abundances of such elements }\citep{kaastra}{. Indeed, in Sect. }\ref{section:ISM_abundances}{ and Table} \ref{table:ISM_oxygen}{ we will validate this assumption.} We obtained a good fit with C-stat/dof \footnote{Here and hereafter \textit{dof} means degrees of freedom.} $=2451/1705$ and $2579/1710$ in the two observations (see Fig. \ref{fig:epic_rgs_fit}). The parameters for both the observations are listed in Table \ref{table:fit}. We designate this simple model where the ISM is modeled with one (neutral) gas component as Model A. The abundances are mostly in agreement between the two observations, but they are not {trustworthy}. In Sect. \ref{sec:rgs_fits} and \ref{section:ISM_abundances} we show that the RGS fit provides a column density higher by 10\%, which significantly changes the abundances estimates. There are also small differences in the continuum parameters, such as the electron temperatures. Indeed we find different temperatures for both comptonization components (see Table \ref{table:fit}). These small deviations affect the broadband spectral slope and forbids to fit the two EPIC-pn observations simultaneously, while this is possible with the RGS spectra.

We also test alternative continuum models in order to show that the adopted model is the best one. \cite{Thompson08} showed that the spectral modeling of GS 1826$-$238 can be done with other continuum models: 1) black body emission plus a single comptonization component; 2) black body emission plus a cut-off powerlaw; 3) double comptonization plus a disk black body; 4) two comptonization components. We test them on both spectra, but report here only the results for the first spectrum. The results for the other observation are similar. The models (1) and (2) give similar results, but with C-stat/dof = {3000/1709} the fit is worse than for our adopted Model A. Model (3) gives even worse fits. The final alternative model (4) gives an intermediate result C-stat/dof = {2625/1707}.

\begin{table}
\caption{EPIC$-$RGS spectral fits to the persistent emission.}

\begin{center}
 \small\addtolength{\tabcolsep}{-2pt}
\scalebox{1}{%
\begin{tabular}{|l|l|l|l|}

\hline
Par / component & OBS 1 & OBS 2 & Average value\\ 
\hline
\hline
{\sl absorber} & & & \\ 
$N_{\rm H}$ \, ($10^{25} \, {\rm m}^{-2}$) &  $ 3.65  \pm 0.05  $  &  $ 3.68  \pm 0.04  $   &  $ 3.67  \pm 0.03  $ \\ 
$kT$        \,     ($10^{-4}$ keV)         &  $ 7.07  \pm 0.10  $  &  $ 7.22  \pm 0.09  $   &  $ 7.15  \pm 0.07  $ \\ 
O          &  $ 1.471 \pm 0.008 $  &  $ 1.403 \pm 0.008 $   &  $ 1.437 \pm 0.006 $ \\ 
Ne         &  $ 2.72  \pm 0.04  $  &  $ 2.64  \pm 0.04  $   &  $ 2.68  \pm 0.03  $ \\ 
Mg         &  $ 0.81  \pm 0.09  $  &  $ 0.80  \pm 0.09  $   &  $ 0.80  \pm 0.06  $ \\ 
Fe         &  $ 1.90  \pm 0.03  $  &  $ 2.01  \pm 0.02  $   &  $ 1.98  \pm 0.02  $ \\ 
\hline                                        
{\sl black body} & & & \\ 
 flux (10$^{-13}$ W m$^{-2}$) & 0.65 $\pm$ 0.10 & 0.55 $\pm$ 0.03 & 0.56 $\pm$ 0.03 \\ 
 $kT_{\rm bb}$ (keV)  &  $ 0.170 \pm 0.002 $      &  $ 0.167 \pm 0.002  $      &  $ 0.168 \pm 0.001 $ \\ 
\hline
{\sl C1 comptonization}       & & & \\ 
 flux (10$^{-13}$ W m$^{-2}$) & 7.8 $\pm$ 0.3 & 7.7 $\pm$ 0.3 & 7.75 $\pm$ 0.21 \\ 
 $kT_{\rm s} $ \hspace{0.22cm}(keV)    &  $ 0.25  \pm 0.02  $      &  $ 0.24 \pm 0.01       $   &  $ 0.24  \pm 0.01  $ \\ 
 $kT_{\rm e} $ \hspace{0.22cm}(keV)    &  $ 2.07  \pm 0.05  $      &  $ 2.38 \pm 0.06       $   &  $ 2.20  \pm 0.04  $ \\ 
 $ \tau $        &  $ 11.0  \pm 0.6   $      &  $ 9.3  \pm 0.4        $   &  $ 9.8  \pm 0.3   $ \\   
\hline
{\sl C2 comptonization}       & & & \\ 
 flux (10$^{-13}$ W m$^{-2}$) & 0.9 $\pm$ 0.1 & 0.6 $\pm$ 0.1 & 0.75 $\pm$ 0.07 \\ 
 $kT_{\rm s} $ \hspace{0.22cm}(keV)    &  $ 0.50  \pm 0.01    $    &  $ 0.54  \pm 0.01    $     &  $ 0.52  \pm 0.01  $ \\ 
 $kT_{\rm e} $ \hspace{0.22cm}(keV)    &  $ 9.4   \pm 0.8     $    &  $ 4.6   \pm 0.5     $     &  $ 5.9   \pm 0.4   $ \\ 
 $ \tau $        &  $\leq 0.5$ &  $\leq 0.7$   &  - \\ 
\hline
C$_{stat}$ / dof & $2451 / 1705$ & $2579 / 1710$ & - \\
\hline

\end{tabular}}
\label{table:fit}
\end{center}
Abundances are relative to the proto-Solar values of \cite{Lodders}. Fluxes are derived in the 0.3-10 keV band. We also report the weighted averages between the two observations. See also Fig. \ref{fig:epic_rgs_fit}.
\end{table} 

\subsection{The high-resolution RGS spectra}
\label{sec:rgs_fits}

\begin{figure*}
\begin{center}
       \includegraphics[bb=10 10 537 694, width=13cm, angle=-90]{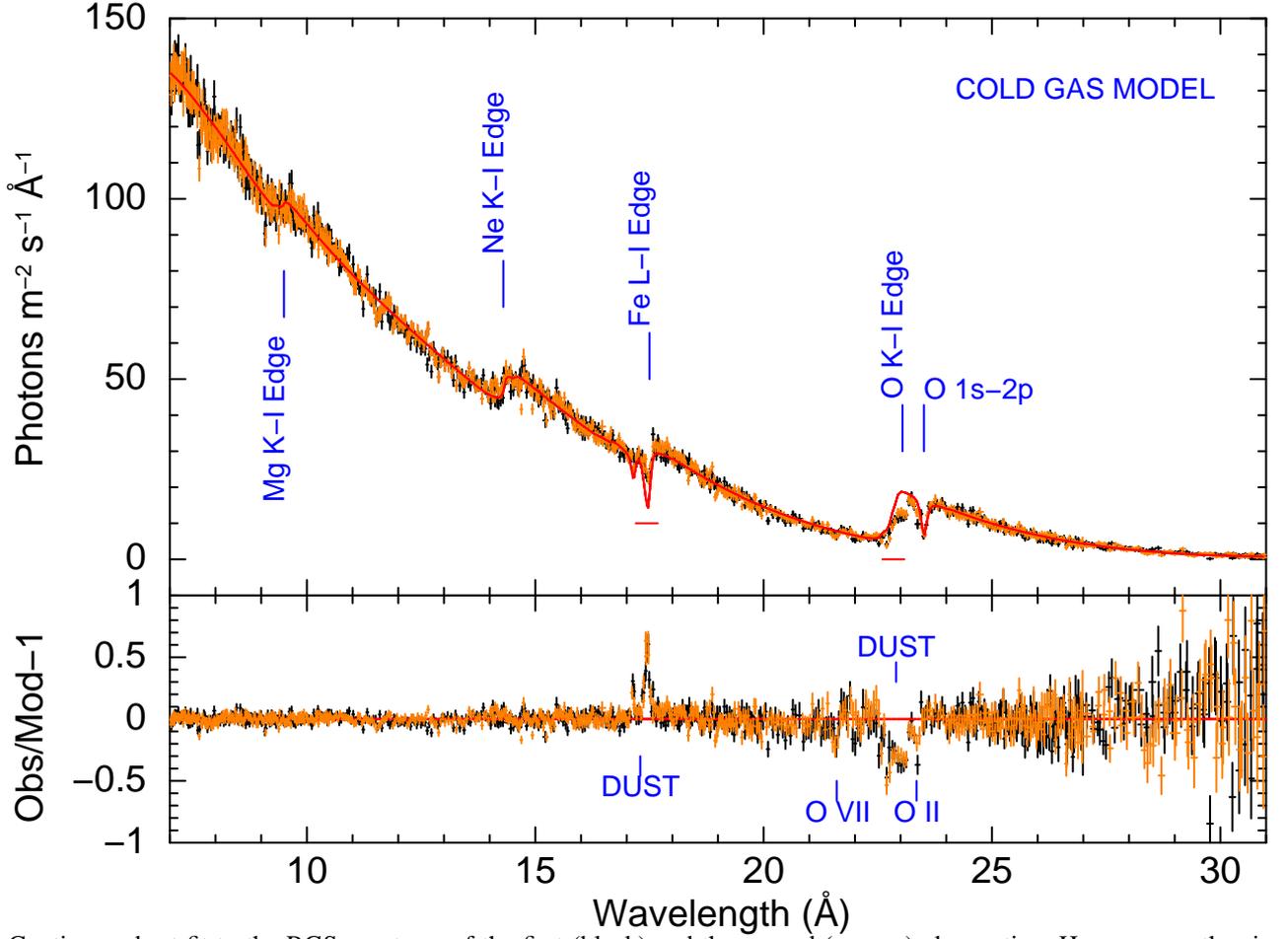}
\end{center}
      \caption{Continuum best fit to the RGS spectrum of the first (black) and the second (orange) observation. Here we use the simple model used for EPIC$-$RGS fits (see Sect. \ref{sec:sim_fits}). In the fits we excluded two small regions near the \ion{O}{i} K-edge and the \ion{Fe}{i} L-edge (see Sect. \ref{sec:rgs_fits}), {which are indicated by two red horizontal strips in the top panel}. See Sect. \ref{sec:dust_analysis} for the detailed analysis. The results of the fits are shown in Table \ref{table:rgs_fit}, they refer to model A.}
          \label{fig:rgs_fit}
    \end{figure*}

In Fig. \ref{fig:rgs_fit} we plot the RGS spectrum of the persistent emission in the first and second observation. Several interesting features can be recognized. At $23.1$ {\AA} we see the absorption edge of the interstellar neutral oxygen, while the \ion{O}{i} line is clearly visible at $23.5$ {\AA}. There is also a broad absorption feature close to the oxygen edge, which is clearly seen in the fits residuals, see Sect. \ref{sec:dust_analysis} for a dedicated discussion. The K-edge of neon and L-edge of iron are easily recognized at $14.3$ {\AA} and $17.5$ {\AA}, respectively. We fit the RGS spectra of the two observations simultaneously with Model A, freezing the shape of the continuum emission and leaving as free parameters the normalizations of the emission components and the parameters of the absorber. In the fits we {still} ignore {the} two small regions ($17.2-17.7$ {\AA} and $22.7-23.2$ {\AA}), close to the iron and oxygen edges respectively (see Sect. \ref{sec:dust_analysis} for the dedicated analysis). The results of the RGS spectral fits are shown in Table \ref{table:rgs_fit} and are designated as Model A. We report also the results of the fits obtained for each observation: the agreement between the parameters validates the simultaneous spectral fit. As expected, the residuals (Fig. \ref{fig:rgs_fit}) show large deviations in the spectral regions that we have temporarily removed: $\sim 4 \sigma$ and $\sim 3 \sigma$ near 17.4 {\AA} and 23 {\AA}, respectively. These features cannot be modeled with a pure-gas model and require the introduction of dust and molecular components in our model (see Sect. \ref{sec:dust_analysis}).

\subsubsection{The neutral gas}
\label{sec:Probe_Cold_gas}

The fits obtained with a simple model (a single gas component for the ISM) show that the ISM can be initially modeled with cold gas (see Table \ref{table:fit} and \ref{table:rgs_fit}, Model A). It has a mean temperature of $kT \sim 6 \times 10^{-4}$ keV{, i.e. about 7$\,$000 K, and provides the bulk of the warm atomic gas anticipated in the introduction}. It is almost neutral, except for iron and magnesium: \ion{Mg}{ii} contributes 30 \% to the total magnesium column density, while \ion{Fe}{ii} accounts for 20 \% of the iron. A precise measure of the ratios \ion{Mg}{i}/\ion{Mg}{ii} and \ion{Fe}{i}/\ion{Fe}{ii} for our spectra is not possible. Near the magnesium edge the spectrum is noisy and the \ion{Mg}{i} and \ion{Mg}{ii} edges are close, at $9.48$ {\AA} and $9.30$ {\AA} respectively, while near the iron edge the lines are unresolved and there is also a contribution from dust that affects the edge structure (see Sect. \ref{sec:dust_analysis}). However, the total magnesium and iron column densities are estimated taking into account the jump across the respective edges and they will not be affected by these problems. As expected, the RGS spectral fits provide a different $N_{\rm H}$ value than the simultaneous EPIC$-$RGS fit, because of the {imperfect EPIC$-$RGS cross-calibration and the low resolution of EPIC, that smooths the absorption features} (see Fig. \ref{fig:epic_rgs_fit}).


\subsubsection{The ionized gas}
\label{sec:Probe_Warm_Hot_gas}

The fit residuals near $21.6$ {\AA} and $23.35$ {\AA} (see Fig. \ref{fig:rgs_fit}), where we should expect the 1s-2p transitions of \ion{O}{vii} and \ion{O}{ii} respectively, suggest the presence of additional weak absorption features. Other weak features are found near $13.4$ {\AA} and $14.6$ {\AA}, close to the theoretical \ion{Ne}{ix} and \ion{Ne}{ii} wavelengths. We deal separately with the different ionization states.

At first we make a fit to the RGS spectra adding columns of \ion{O}{ii} and \ion{Ne}{ii} to our model through a {\sl slab} component. The {\sl slab} model calculates the transmission of a layer of plasma with arbitrary composition. Free parameters are the intrinsic velocity dispersion and the column densities of the individual ions \citep{kaastraspex}. The fits improve significantly: {by fitting the two RGS observations simultaneously we get $\Delta$C-stat $\sim$ 130}. The velocity dispersion is not well constrained ($\sigma_V$ = 50 $\pm$ 15 km s$^{-1}$). The average ion columns are {1.2} $\pm$ 0.4 $\times 10^{21}\, {\rm m}^{-2}$ (\ion{O}{ii}) and {2.4} $\pm$ 0.4 $\times 10^{21}\, {\rm m}^{-2}$ (\ion{Ne}{ii}), while the cold gas gives 3.05 $\pm$ 0.15 $\times 10^{22} \,{\rm m}^{-2}$ (\ion{O}{i}) and 6.7 $\pm$ 0.3 $\times 10^{21} \,{\rm m}^{-2}$ (\ion{Ne}{i}).

In second instance we add another {\sl slab} component to take into account the contribution by the hot gas. The columns are {1.1 $\pm$ 0.5} $\times 10^{20}\, {\rm m}^{-2}$ (both \ion{O}{vii} and \ion{O}{viii}) and {3.5 $\pm$ 2.5} $\times 10^{19} \,{\rm m}^{-2}$ (\ion{Ne}{ix}). {The addition of the hot gas provides $\Delta$C-stat $= 30$, which is significantly less than the improvement we have obtained by adding the low-ionized gas.} Moreover we can only put an upper limit to the velocity dispersion of the hot ionized gas (250 km s$^{-1}$).

\begin{figure}
    \centering
      \includegraphics[angle=-90, width=9cm]{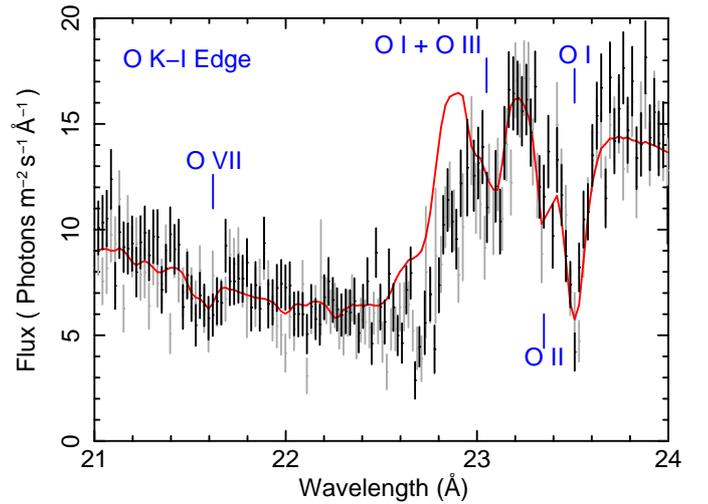}
\caption{\ion{O}{i} K-edge: data and Model B. Black and grey points refer to the first and second observation, respectively.}%
    \label{fig:rgs_edge_Ox}
    \end{figure}

\begin{figure}
    \centering
      \includegraphics[angle=-90, width=9cm]{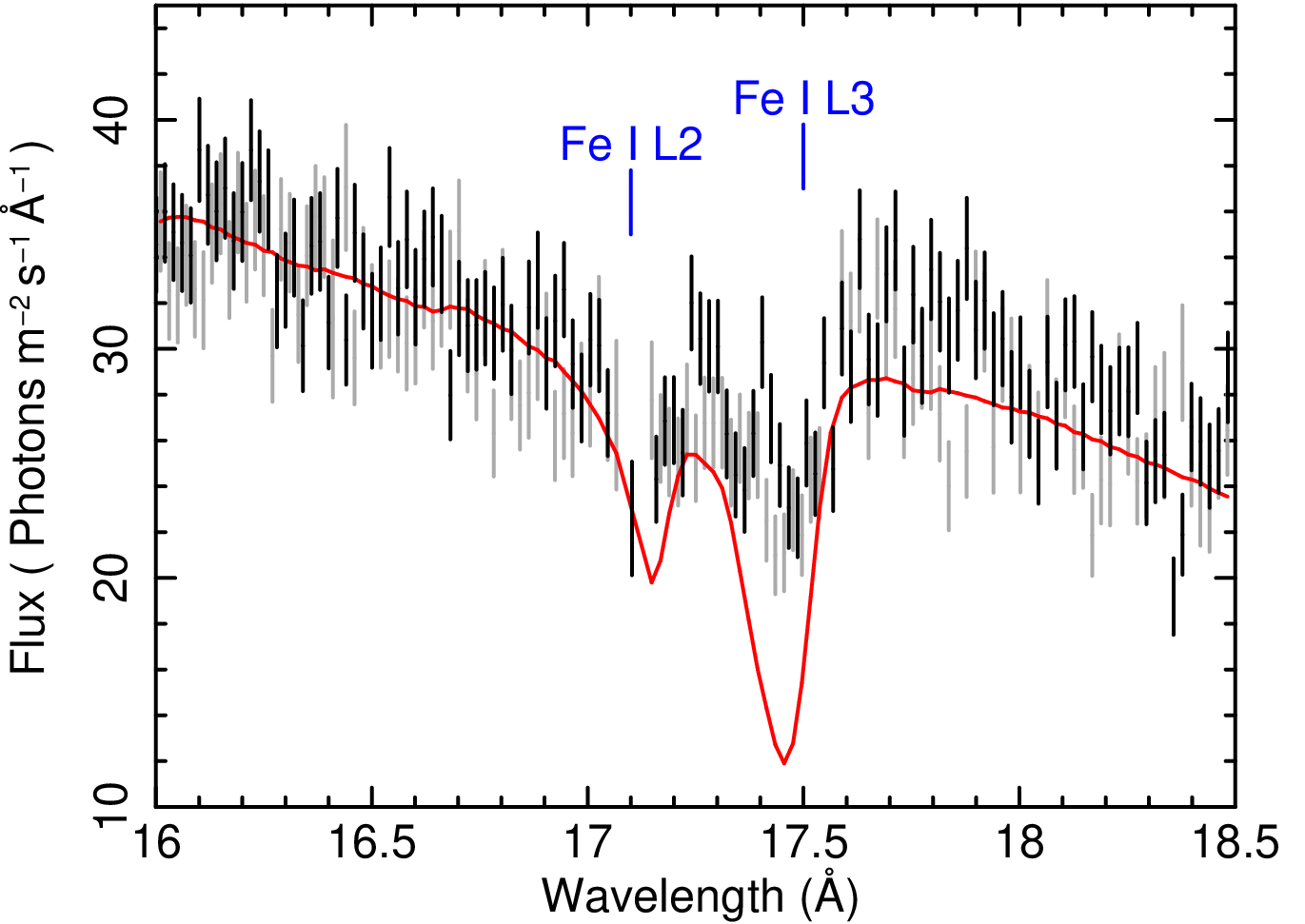}
\caption{\ion{Fe}{i} L-edge: data and Model B.}%
    \label{fig:rgs_edge_Fe}
    \end{figure}

\begin{figure}
    \centering
      \includegraphics[angle=-90, width=9cm]{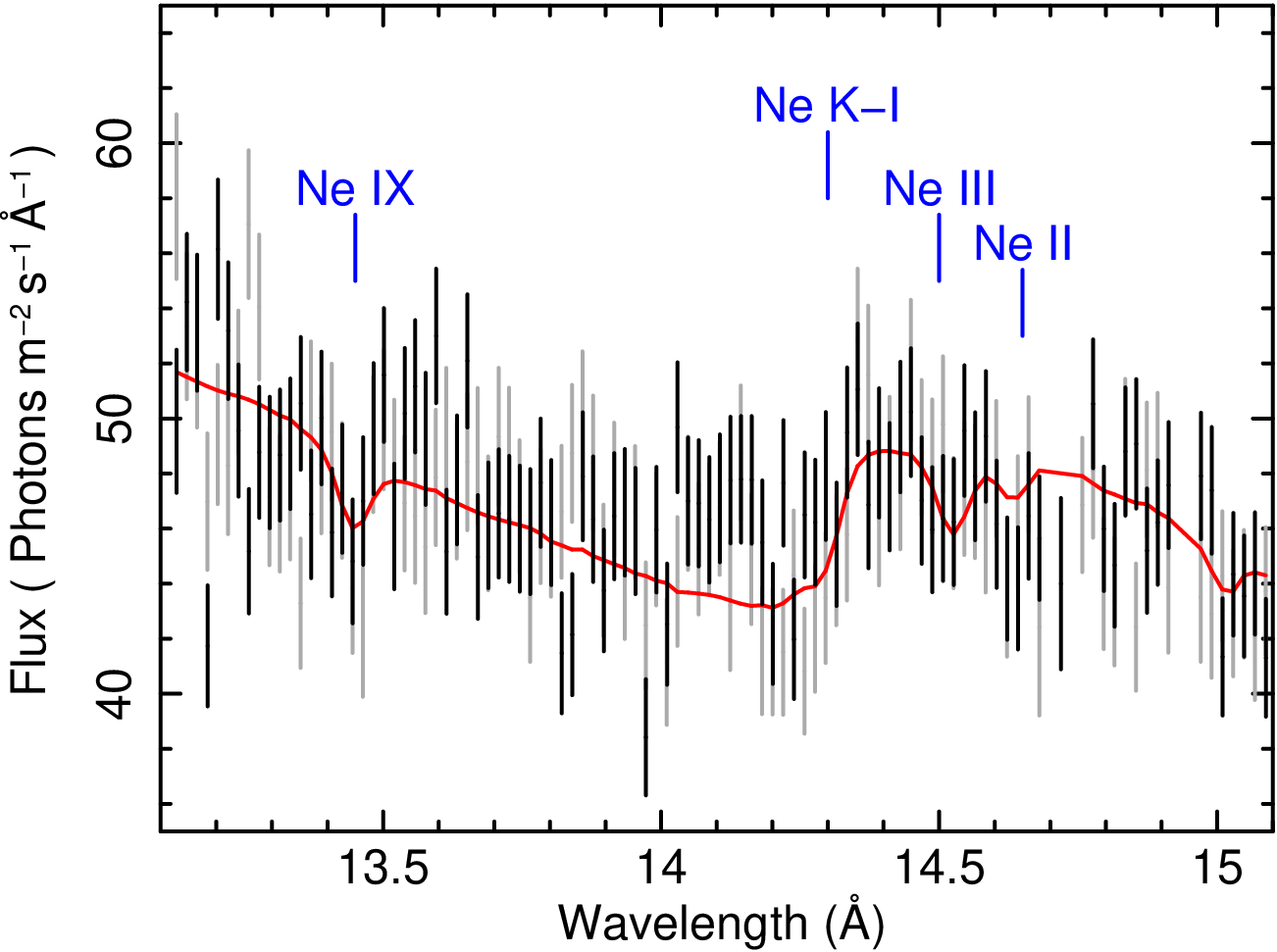}
\caption{\ion{Ne}{i} K-edge: data and Model B.}%
    \label{fig:rgs_edge_Ne}
    \end{figure}

\begin{figure}
    \centering
      \includegraphics[angle=-90, width=9cm]{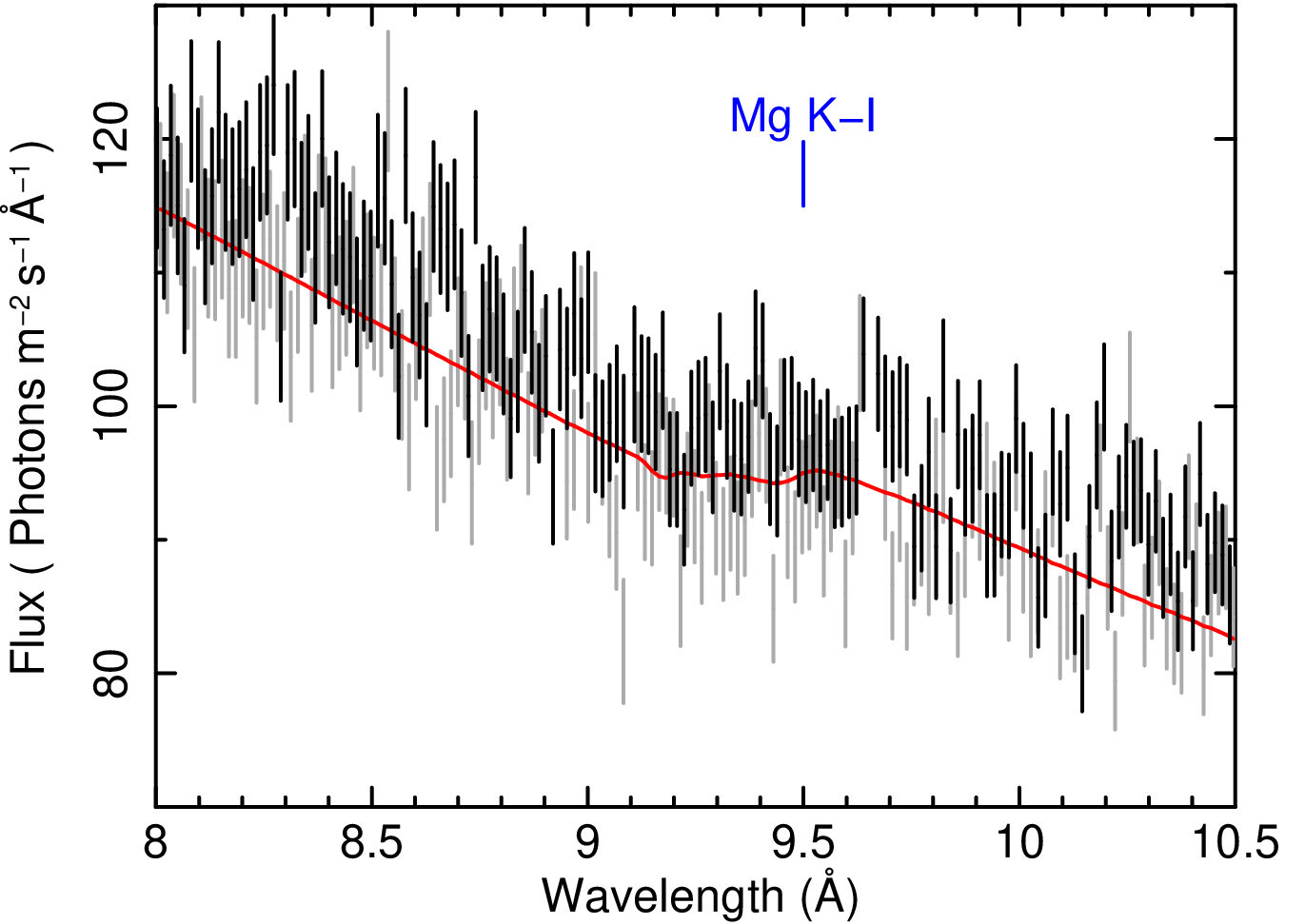}
\caption{\ion{Mg}{i} K-edge: data and Model B}%
    \label{fig:rgs_edge_Mg}
    \end{figure}

However, in order to take care of every absorption feature created by all ions in the warm-hot phases and to deal with physical models, we substitute the two {\sl slab} components with two {\sl hot} components. We couple the elemental abundances of the warm-hot components to those of the cold gas, assuming all ISM phases have the same abundances. This is a reasonable assumption, especially for the warm (low-ionization) ionized gas, as its temperature is not too different from the temperature of warm neutral gas. The additional free parameters are the hydrogen column density, the temperature and the velocity dispersion. In summary, the additional warm and hot phases give an average improvement of $\Delta$C-stat $\sim 80$ for only 6 free parameters added. We label such a 3-gas model as Model B and display all results in Table \ref{table:rgs_fit}. We plot the individual absorption edges of O, Fe, Ne and Mg in Fig. \ref{fig:rgs_edge_Ox}, \ref{fig:rgs_edge_Fe}, \ref{fig:rgs_edge_Ne} and \ref{fig:rgs_edge_Mg}, respectively. We discuss these results in Sect. \ref{sec:ISM_structure}. The predicted deviations near the oxygen and iron edges, clearly seen in Fig. \ref{fig:rgs_edge_Ox} and \ref{fig:rgs_edge_Fe}, still confirm that pure interstellar gas cannot reproduce all the absorption features and that we need to take into account different states of matter, such as solids.

Finally, we also consider the model where the cold gas component is forced to be neutral by freezing its temperature to $5\times10^{-4}$ keV, i.e. $5\,800$ K. In this case, we get an almost equal fit (C-stat/dof = {4466/3227}), with the exception that the warm component {has a significantly} lower temperature ($\sim 1.4\times10^{-3}$ keV, i.e. $10-20\,000$ K), to account for the \ion{O}{ii} that in our nominal fit is partially produced by the cold component. {This temperature value is more representative than our previous value of $5.4\times10^{-3}$ keV for the warm ionized gas in the ISM} \citep{ferriere}. However, both the fits are acceptable, thus we report only results obtained with the cold-gas temperature as free parameter in Table \ref{table:rgs_fit}.

\begin{table*}
\caption{RGS spectral fits to the persistent emission.}
\begin{center}
\scalebox{1}{%
\begin{tabular}{|l|l|l|l|l|l|l|l|}
\hline
                      &                                        & \multicolumn{3}{c|}{Mod A $^{(a)}$}   & Mod B & Mod C \\
\hline
Component             & Parameter                              & OBS 1         & OBS 2         & OBS 1+2        & OBS 1+2         & OBS 1+2 \\ \hline
\multirow{7}{*}{Cold} & $N_{\rm H} \, (10^{25} {\rm m}^{-2})$  & $4.18\pm0.01$ & $4.22\pm0.07$ & $4.21\pm0.09$  & $3.88\pm0.07$   & $3.94\pm0.05$ \\
                      & $kT$ ($10^{-4}$ keV)                   & $5.90\pm0.14$ & $6.13\pm0.12$ & $6.02\pm0.09$  & $6.04\pm0.10$   & $8.6\pm0.4$ \\
                      & $\sigma_V$ (km s$^{-1}$)               & $27\pm15$     & $18\pm6$      & $13\pm7$       & $< 12.6$        & $< 24$         \\
                      & O                                      & $1.30\pm0.02$ & $1.29\pm0.02$ & $1.29\pm0.01$  & $1.29\pm0.02$   & $1.17\pm0.03$\\
                      & Ne                                     & $2.08\pm0.04$ & $1.86\pm0.07$ & $1.95\pm0.07$  & $2.19\pm0.10$   & $1.75\pm0.11$\\
                      & Mg                                     & $2.27\pm0.12$ & $2.14\pm0.16$ & $2.21\pm0.16$  & $1.93\pm0.15$   & $1.30\pm0.25$\\
                      & Fe                                     & $1.39\pm0.02$ & $1.42\pm0.06$ & $1.39\pm0.05$  & $1.65\pm0.08$   & $ < 0.1 $\\
\hline
\multirow{3}{*}{Warm} & $N_{\rm H} \, (10^{25} {\rm m}^{-2})$  & -             & -             & -              & $0.46 \pm 0.06$ & $0.15 \pm 0.05$\\
                      & $kT$ ($10^{-3}$ keV)                   &    -          &     -         & -              & $5.4\pm0.3$     & $4.5\pm0.5$ \\
                      & $\sigma_V$ (km s$^{-1}$)               & -             &   -           &  -             & $50\pm25$       & $< 150$ \\
\hline 
\multirow{3}{*}{Hot}  & $N_{\rm H} \, (10^{25} {\rm m}^{-2})$  & -             & -             &   -            & $0.042\pm0.008$ & $0.047\pm0.011$\\
                      & $kT$ (keV)                             &  -            & -             &    -           & $ 0.20\pm0.02 $ & $0.20\pm0.03$\\
                      & $\sigma_V$ (km s$^{-1}$)               & -             & -             &     -          & $< 160$         & $< 200$ \\
\hline
\multirow{2}{*}{Dabs $^{(b)}$} & $N_{\rm Fe}$ & - & - & - & - & $2.6\pm0.1$ \\
                               & $N_{\rm Mg}$ & - & - & - & - & $2.3\pm0.1$ \\
\hline
\multirow{4}{*}{Amol $^{(b,\,c)}$} & $N_{\rm O}$ (i=14, Silicates)  & - & - & - & - & $2.5\pm0.5$ \\
                               & $N_{\rm O}$ (i=7, H$_2$O Ice)        & - & - & - & - & $< 0.7$ \\
                               & $N_{\rm O}$ (i=2, CO)         & - & - & - & - & $< 0.4$ \\
                               & $N_{\rm O}$ (i=23, Aluminates) & - & - & - & - & $< 0.4$ \\
\hline
\multirow{2}{*}{Statistics} & C-Stat / dof     & $2239/1595$   & $2170/1589$ & $4587/3232$ & $4435/3226$  & {4818/3398} \\
                            & C-Stat / dof (*) & $2930/1684$   & $3040/1710$ & $5757/3410$ & $6064/3404$  & {4818/3398} \\
\hline
\end{tabular}}
\label{table:rgs_fit}
\end{center}
$^{(a)}$ We give the separate fits for the two observations only for Model A in order to show that they are consistent within the errors and thus can be fitted together. C-Stat / dof (*) refers to the statistics obtained by including the wavelength ranges 17.2-17.7 {\AA} and 22.7-23.2 {\AA}, which in fits are ignored except in the case of the complete model. $^{(b)}$ All the columns for the {\sl dabs} and {\sl amol} components are reported in units of $10^{21} {\rm m}^{-2}$. $^{(c)}$ Each {\sl amol} component is displayed together with its molecular index as reported in Table \ref{tab:amol}.
\end{table*} 

\subsubsection{Fine structures: Dust and molecules}
\label{sec:dust_analysis}

\begin{figure}
      \centering
      \includegraphics[angle=-90, width=9cm]{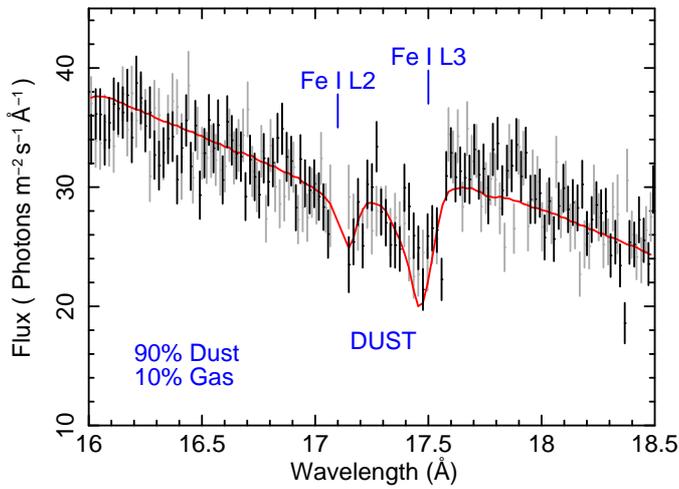}
      \caption{Iron edge: data and Model C.}
          \label{fig:iron_fine_edge}
    \end{figure}

Further important improvements to our fit are obtained by adding both dust and molecules to our multiphase gas model. We use here two recently added models of SPEX: {\sl dabs} and {\sl amol}, which are described briefly below. The transmission of dust is calculated by the {\sl dabs} model in SPEX: it accounts for the self-shielding of X-ray photons by dust grains, but uses the edge and line structure for atomic gas. It was first used in the analysis of the Crab spectrum by \citet{kaastra}. It follows completely the dust treatment as described by \cite{Wilms} and is useful to estimate the dust-to-gas column ratio and the depletion factor for several elements. We assume the default values for the grain parameters because they are physically acceptable: the grains are assumed to be spherical and fluffy, with density $\rho = 1000$ kg m$^{-3}$, grain radius $a$ between $a_{\rm min} \, < \, a < \, a_{\rm max}$, where $a_{\rm min} = 0.025 \, \mu$m and $a_{\rm max} = 0.25 \, \mu$m, with a size distribution $dn/da \sim a^{-p}$ and $p=3.5$ \citep{kaastra}. Including the {\sl dabs} component in our model, we obtain a significant improvement to the fit by requiring at least $\sim$ 90\% of iron to be confined in dust grains (see Table \ref{table:rgs_fit}). Inside the iron edge the shielding effect of dust is stronger than its fine structure features and thus the {\sl dabs} component is suitable to fit the data. Indeed, the $\sim$ 4 $\sigma$ deviation at $17.4$ {\AA} (Fig. \ref{fig:rgs_edge_Fe}) and the $\sim 1 \sigma$ deviation near $17.1$ {\AA}, due to the assumption of a pure-gas ISM, just disappear (see Fig. \ref{fig:iron_fine_edge}). Moreover, from the {\sl dabs} model we derive $\sim$ 40\% of the oxygen to be bound in dust grains. Unfortunately, this model does not yield a good fit of the oxygen edge, where strong features due to molecules are present that are not taken into account by the {\sl dabs} model.

\begin{figure}
      \centering
      \includegraphics[angle=-90, width=9cm]{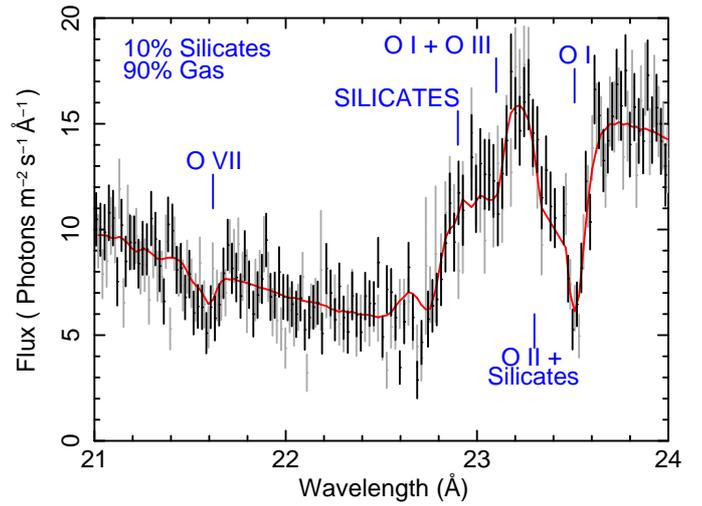}
      \caption{{Oxygen edge: data and Model C.}}
          \label{fig:oxygen_fine_edge}
    \end{figure}

A much larger improvement is obtained when we introduce molecules and minerals containing oxygen atoms. For the first time we use the {\sl amol} model in SPEX to take into account bound forms of oxygen. The model currently takes into account the modified edge and line structure around the O K-edge using measured cross sections of various compounds, taken from the literature. More details about this model can be found in Appendix \ref{appendix:amol}. The {\sl amol} model is very useful to constrain the local molecular features, but it does not account for the dust shielding effects. We have tried 23 different types of compounds such as CO, N$_2$O, H$_2$O, ice, FeO and several minerals. The best fit is obtained by using a mixture of silicates, such as andradite, and water ice plus other molecules (see Fig. \ref{fig:oxygen_fine_edge} and Table \ref{table:molecules}). This additional component removes the previous 3 $\sigma$ deviation of the pure-gas model between 22.7-23.0 {\AA} inside the oxygen edge (Fig. \ref{fig:rgs_edge_Ox}). Finally we complete the dust model choosing the {\sl amol} component for oxygen and {\sl dabs} component for all the other depleted elements, such as iron and magnesium. The final gas+dust model describes the data much better than all previous models (see Fig. \ref{fig:iron_fine_edge} and \ref{fig:oxygen_fine_edge}) and it allows to estimate the dust-to-gas ratio. We label this model C and show the parameters in Table \ref{table:rgs_fit}.

\subsection{ISM model complexity}
\label{sec:ISM_structure}

The spectral modeling indicates that in the line of sight towards our X-ray source the ISM is much more complex than simple neutral gas. The gas is rather structured in different phases and dust also consists of various compounds.

The gas consists of 3 components (see Table \ref{table:rgs_fit}): cold gas with a temperature $kT \sim 5-10 \times 10^{-4}$ keV {(5.8$-$10 $\times 10^3$ K)}, warm ionized gas with $kT \sim 1-5 \times 10^{-3}$ keV {(1$-$6 $\times 10^4$ K)} and hot ionized gas with $kT \sim 0.2$ keV {($\gtrsim 2 \times 10^6$ K)}. The column densities $N_{\rm H}$ of these three components span over 2 orders of magnitude: the cold gas accounts for $\sim 90-95$ \% of the total column $N_{\rm H}^{\rm tot}$, $N_{\rm H}^{\rm warm} \sim 5-10$  \% of $N_{\rm H}^{\rm tot}$, while the hot gas contributes $\sim 1$ \%.

The warm component produces the low-ionization absorption lines of \ion{O}{ii} and \ion{O}{iii}, at $23.35$ {\AA} and $23.1$ {\AA} respectively (see Fig. \ref{fig:rgs_edge_Ox}). It also provides a better modeling of the neon edge (see Fig. \ref{fig:rgs_edge_Ne}). Using model B we estimate N$_{\ion{O}{ii}} = 2.0 \pm 0.5 \times 10^{21} \, {\rm m}^{-2}$ and N$_{\ion{O}{iii}} = 1.4 \pm 0.5 \times 10^{21} \, {\rm m}^{-2}$, respectively $\sim 7\%$ and $\sim 5\%$ of the total oxygen column. However, the derived column density of the warm ionized gas is affected by the presence of dust and molecules on the line of sight. Indeed, the absorption features that we see near $23.35$ {\AA} and $23.1$ {\AA} are contaminated by dust and molecules effects. Contributions from dust and molecules (Model C in Table \ref{table:rgs_fit}) are confirmed by the improvements to the fit (see also Fig. \ref{fig:iron_fine_edge} and \ref{fig:oxygen_fine_edge}) and the column density of the warm gas is finally reduced to about 5\% of the full gas column (see also Table \ref{table:molecules}).

The column density of hot gas is about two orders of magnitude lower than the cold gas column and its temperature is around two million K. As expected, the hotter gas has a higher velocity dispersion (see Table \ref{table:rgs_fit}). The hot gas model gives a good fit of the \ion{O}{vii} absorption line at $21.6$ {\AA}, together with the small feature at $13.4$ {\AA} produced by \ion{Ne}{ix}.

According to the analysis of the oxygen edge, the solid phase of the ISM towards GS 1826$-$238 consists of a mixture of minerals (such as andradite silicates) and traces of CO and water ice. We cannot yet distinguish between amorphous and crystalline phases. As reported in Table \ref{table:molecules} the bulk of the oxygen, $\sim 90 \%$, appears to be in the gas phase, while the remaining $\sim 10 \%$ is made mostly of solids, such as silicates and water ice. Obviously, there could be substances able to reproduce such features in the spectrum other than our few dozen test molecules. For the iron, instead, we obtain a different composition: at least $\sim 90 \%$ of Fe appears to be bound in dust grains. In our dust model we assume a depletion factor of 0.8 for magnesium, as suggested by \cite{Wilms}. The derived gas ($2.0 \pm 0.4 \times 10^{21}\,{\rm m}^{-2}$) and dust ($2.3 \pm 0.1 \times 10^{21}\,{\rm m}^{-2}$) column densities for the Mg are identical within the errors (see also Table \ref{table:rgs_fit}).

\begin{table}
\caption{Table of the contributions to the oxygen column-density.}  
\begin{center}
 \small\addtolength{\tabcolsep}{-2pt}
\scalebox{1}{%
\begin{tabular}{|l|l|l|l|l|}
\hline
Phase               & Constituent & $N_{\rm O}\,(10^{22}$ m$^{-2})$ & \% of $N_{\rm O}^{p}\,^{(a)}$& \% of $N_{\rm O}\,^{(b)}$\\
\hline
\multirow{3}{*}{Gas}       & \ion{O}{i}                             & $2.7 \pm 0.1 $  & $94 \pm 4 $    &   \\
                           & \ion{O}{ii}, \ion{O}{iii}, \ion{O}{iv} & $0.10 \pm 0.05$ & $4 \pm 2 $     & $90 \pm 6 $\\
                           & \ion{O}{vii}, \ion{O}{viii}            & $0.05 \pm 0.01$ & $2.0 \pm 0.5 $ & \\
\hline
\multirow{2}{*}{Dust}      & Silicates                              & $0.25 \pm 0.05$ & $85-100$  & \multirow{2}{*}{$10\pm 2$} \\
                           & Aluminates                             & $< 0.04$        & $0-15$    &    \\
\hline
\multirow{2}{*}{Molecules} & H$_2$O ice                             & $< 0.07$        & $\sim 65$ & \multirow{2}{*}{$0-2$}  \\
                           & CO                                     & $< 0.04$        & $\sim 35$ &                         \\
\hline
\end{tabular}}
\end{center}
$^{(a)}$ \% of $N_{\rm O}^{p}$ represents the contribution of each constituent to the respective phase. $^{(b)}$ \% of $N_{\rm O}$ give the contribution of the different phases to the total oxygen column density. See also Sect. \ref{sec:dust_analysis} and Fig. \ref{fig:oxygen_transmission} for the transmission of the main compounds.
\label{table:molecules}
\end{table} 

    \begin{figure}
      \centering
      \includegraphics[width=9cm]{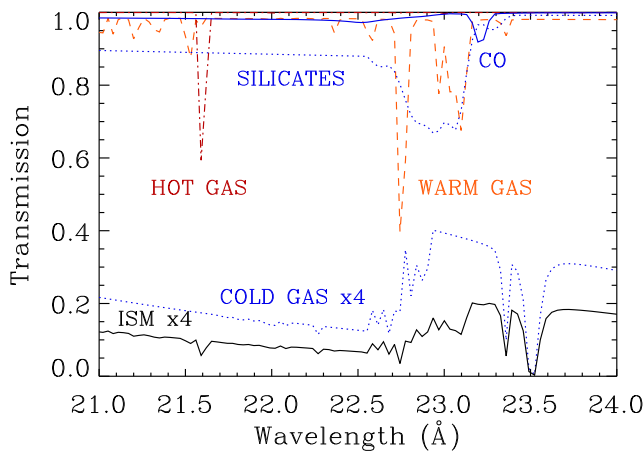}
      \caption{Transmission near the \ion{O}{i} K-edge. The model used is Mod C (see Sect. \ref{sec:dust_analysis}, Table \ref{table:rgs_fit} and \ref{table:ISM_oxygen}). Both cold-gas and entire-ISM transmission is multiplied by a factor of 4 for displaying purpose.}
          \label{fig:oxygen_transmission}
    \end{figure}

\subsection{ISM abundances}
\label{section:ISM_abundances}

We have estimated the abundances of Mg, Ne, Fe, O and N (Table \ref{table:ISM_abundances}). The column density of each element refers to the sum of the contributions from all the gas and dust phases. The abundances do not differ significantly between the two observations (see Table \ref{table:rgs_fit}). This is expected if the absorption is mainly due to the interstellar medium, because the ISM is stable on short time scales. The zero shift of the \ion{O}{i} line, the position of the O, Fe, Ne and Mg edges (see Fig. \ref{fig:rgs_edge_Ox} to  \ref{fig:oxygen_fine_edge}) and the low velocity dispersion suggest that the absorber matter is a mixture of gas and dust without outflows or inflows, not broadened due to Keplerian motion around the X-ray source. This is also consistent with an ISM origin. For further discussion on the abundances and comparisons with previous work see Sect. \ref{section:discuss_compare}.

A separate analysis is required for the nitrogen edge at 30.8 {\AA}. Indeed, at wavelengths higher than 29 {\AA} the source flux decreases significantly and is comparable to the background. Thus, the spectrum around the \ion{N}{i} K-edge is noisy and the \ion{N}{i} column density has a large uncertainty. We have decided to freeze the continuum parameters and try either a broadband fit to the range 7-33 {\AA} or a local fit between 27-33 {\AA}. The abundance estimates are consistent within the error and their average is reported in Table \ref{table:ISM_abundances}.

\begin{table*}
\caption{Average ISM abundances in units of the proto-Solar values \citep{Lodders} and calculated by summing contribution from all ISM phases.}             
\begin{center}
\scalebox{1}{%
\begin{tabular}{|l|l|l|l|l|l|}
\hline
X                      &  O                & Ne             & Mg            & Fe            & N $^{(e)}$       \\
\hline
\object{GS 1826-238} $^{(a)}$   &  $1.23\pm0.05$    & $1.75\pm0.11$  & $2.45\pm0.35$ & $1.37\pm0.17$ & $2.4\pm0.7$      \\
Crab $^{(b)}$          &  $1.030\pm0.016$  & $1.72\pm0.11$  & $0.85\pm0.21$ & $0.78\pm0.05$ & $1.01\pm0.09$    \\
Cyg X-2 $^{(c)}$       &  $0.6-0.8$        & $0.8-1.1$      & $0.6-1.1$     & -             & -                \\
\object{4U 1820-303} $^{(d)}$   &  $0.7-1.1$        & $1.1-2.0$      & -             & $0.3-0.8$     & -                \\
\hline
\end{tabular}}
\end{center}
$^{(a)}$ Model C, gas + dust. $^{(b)}$ \cite{kaastra} Model B, gas + dust. $^{(c)}$ \cite{yao09}. $^{(d)}$ \cite{JuettII}. \\ $^{(e)}$ Estimated through a local fit in the 27-33 {\AA} range.
\label{table:ISM_abundances}
\end{table*}

\begin{table}
\caption{Ionic column densities of oxygen in $10^{22}$ m$^{-2}$.}             
\begin{center}
\scalebox{1}{%
\begin{tabular}{|l|l|l|l|l|}
\hline
X                                       & Mod A$_{sim}$ $^{(a)}$  & Mod A $^{(a)}$  & Mod B $^{(b)}$  & Mod C $^{(c)}$ \\ 
\hline
\ion{O}{i}                              & 3.0                     &  3.1            &  2.9            & 2.7 \\
\ion{O}{ii}, \ion{O}{iii}, \ion{O}{iv}  & $\equiv 0$                       &  $\equiv 0$             &  0.35           & 0.1 \\
\ion{O}{vii}, \ion{O}{viii}             & $\equiv 0$                      &  $\equiv 0$              &  0.05           & 0.05 \\
Dust                                    & $\equiv 0$                       &  $\equiv 0$              &  $\equiv 0$              & 0.35 \\
\hline
Total $N_{\rm O}$                       & 3.0                     &  3.1            &  3.3            & 3.2 \\
\hline
\end{tabular}}
\end{center}
$^{(a)}$ Mono-phase gas (Table \ref{table:rgs_fit} for the simultaneous EPIC-RGS fit). $^{(b)}$ Three-phases gas (Table \ref{table:rgs_fit}). $^{(c)}$ Gas + dust model (see Table \ref{table:molecules}). {The agreement between the oxygen column densities estimated with different models validates our method.}
\label{table:ISM_oxygen}
\end{table}

\section{Discussion}
\label{discussion}

\subsection{The continuum}
\label{sec:continuum}

Our analysis shows that the persistent state of the low-mass X-ray binary GS 1826$-$238 is well represented by a double comptonization (C1+C2) plus a black body (BB) emission component, all three absorbed by the interstellar medium composed of a 3-phases gas, dust and molecules (see Table \ref{table:fit} and \ref{table:rgs_fit}). In both observations the fits are in agreement: all the ISM parameters appear to be fully consistent, thus we can discuss about the results from the simultaneous fit of the high-resolution RGS data.

\subsection{ISM structure}

In Sect. \ref{sec:ISM_structure} we show that in our line of sight the ISM has a clear multiphase structure. There are media with different ionization states, dust grains and molecules. As confirmed by Fig. \ref{fig:oxygen_transmission}, the bulk of the matter responsible for X-ray absorption is found in the form of \textsl{cold} gas with a temperature $\sim$ 7$\,$000 K and low velocity dispersion ($\sigma_V \lesssim 13$ km s$^{-1}$). At this temperature the gas is almost neutral: only iron and magnesium are partially ionized. Part of the cold matter is found in solids compounds, such as \textsl{dust grains} and \textsl{molecules}. Most of iron is bound in dust grains. About 10 \% of oxygen is found in compounds: the silicates contribute to $\sim$ 80\% of this phase, while the remaining fraction consists of a mixture of other oxides (such as iron aluminates) together with ices and CO molecules in similar quantities (see Table \ref{table:molecules}). The best fit is obtained using as compound the andradite silicate Ca$_3\,$Fe$_2\,$(SiO$_4$)$_3$, but we need higher signal-to-noise data to distinguish among the different silicates, as also olivine and pyroxene are good candidates. Moreover, at the present stage our model does not take into account simultaneously the shielding and fine structure effects of oxygen compounds, thus the fraction of oxygen bound in solids could be higher, e.g. up to 40 \% (see Sect. \ref{sec:dust_analysis}). However we are working to the development of models that take into account all the possible effects and we postpone a deeper analysis of the oxygen dust phase to a forthcoming paper.

About 5\% of the gas is ionized (see Mod C in Table \ref{table:rgs_fit}). Most is a \textsl{warm} plasma with $T \sim 10-50\,000$ K. It has a low-ionization degree and it accounts for the interstellar \ion{O}{ii} and \ion{O}{iii}. Only 1-2\% of the ISM gas appears to be highly ionized. Such a \textsl{hot} plasma reaches temperatures $\sim 2 \times 10^{6}$ K and contributes to all \ion{O}{vii} and \ion{O}{viii} present in the ISM. As we expect, the higher the temperature of the gas phase, the higher its velocity dispersion. Unfortunately the velocity dispersion is not well constrained, especially for the hot ionized gas. This is not surprising as the lines are unresolved.

\subsection{Comparison with previous results}
\label{section:discuss_compare}


\subsubsection{ISM constituents}
\label{section:discuss_constituents}

The average total column density of the multi-phase gas we estimate is about $(4.14 \pm 0.07) \times 10^{25}\, $m$^{-2}$. This is not consistent with the value of $(3.19 \pm 0.01) \times 10^{25} \, $m$^{-2}$ found by \citet{Thompson08}. They combined data from Chandra, \textit{XMM-Newton} and RXTE in the $0.5-100$ keV band, but we know that the bulk of the absorption is at lower energy. Instead we have used both EPIC and RGS data. The latter detector has higher spectral resolution between $0.3-2$ keV and allows to better estimate the absorption column. Also, \citet{Thompson08} used the \citet{Morrison} model to fit absorbing medium, which takes into account only contribution by cold neutral gas. In Sect. \ref{sec:sim_fits} we have shown that such a model is not optimal, because it does not include lines, then abundances and temperature cannot be free parameters. Instead our estimate of $N_{\rm H}$ is obtained by summing contribution from all the phases of the gas and by accounting for all the absorption features found in the spectrum.

The multiphase structure of the ISM that we have constrained is consistent with recent results \citep{ferriere, yao06, yao09}. First of all, there is a good agreement in the fractions between the cold, warm and hot phases of the gas. In particular the estimated amount of \ion{O}{vii}, $\sim 1.6 \times 10^{20} \, {\rm m}^{-2}$, is fully consistent with the value found by \cite{yao06} by fitting both \ion{O}{vii} 1s$-$2p and 1s$-$3p lines in the spectrum of \object{4U 1820$-$303}, which is another LMXB near the center of the Galaxy. The hot gas accounts for $1-2$ \% of the total column density and it represents the average fraction of hot plasma in the Galaxy. Indeed it agrees with previous estimates and can be fully explained by the heating of supernovae \citep{McCammon}. The velocity dispersion $\sigma_V$ estimates agree with recent work. \citet{yao06} combined oxygen and neon ionization lines in the {\sl Chandra} spectrum of the LMXB \object{4U 1820$-$303} obtaining $\sigma_V < 350$ km s$^{-1}$. \citet{JuettI} found $\sigma_V < 200$ km s$^{-1}$ from oxygen lines fits to the {\sl Chandra} spectra of several LMXBs.

Moreover, we found clear indications of dust depletion in some heavy elements, such as oxygen, iron and magnesium. According to the different dust models that we have used ({\sl dabs} and {\sl dabs+amol}), we find $\sim$ 50\% of Mg, more than 90\% of Fe and $10-40$\% of oxygen in the form of dust grains and molecular compounds. Such results are mostly consistent with previous estimates \citep{Wilms, kaastra}. The iron dust-to-gas ratio found towards GS 1826$-$238 is among the highest measured in the Galaxy. \citet{Williams96} showed that the higher is the density of a molecular cloud, the higher is the probability of forming dust and molecules from gas particles. Thus the higher dust-to-gas ratio we estimated towards the center of the Galaxy suggests high density regions.

The presence of silicates and ice in the ISM, constrained by the \ion{O}{i} K-edge analysis, is supported by other work. \cite{paerels} found similar features within $22.7-23.0$ {\AA} in the spectrum of the LMXB \object{X0614+091} and they argued that it should be due to iron oxides or oxygen generally bound in dust. \citet{costantini2005} found indications of silicates such as olivine and pyroxene by modeling the feature of the scattering halo of Cyg X$-$2. Recently, \cite{costantini} found evidence for EXAFS in the short-wavelength side of the oxygen edge in the spectrum of Sco X$-$1 and their results suggest the presence of amorphous water ice.

We have checked CO surveys \citep{Dame_CO} in order to test the presence of molecular clouds. There is no clear evidence for these clouds in our line of sight $(x,y)_{GAL} \sim (9.3^{\circ},-6.1^{\circ})$, while there is an important amount of diffuse dust \citep[{see also}][]{Schlegel98}. This is consistent with the fact that we can only put upper limits to ice and CO columns. Thus the absorption should arise from a solid phase consisting mostly of minerals.

    \begin{figure}
      \centering
      \includegraphics[width=9cm]{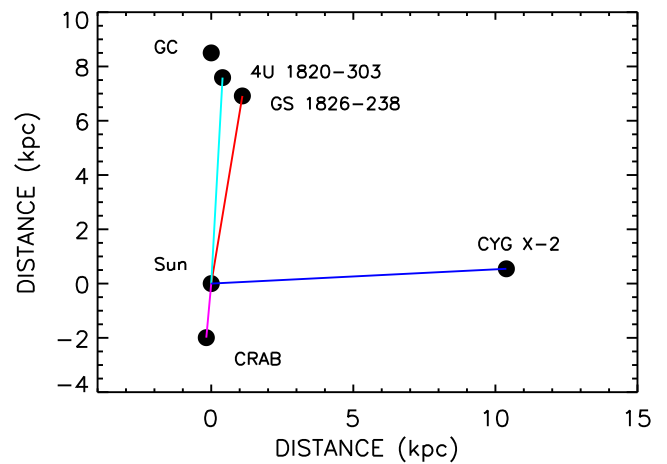}
      \caption{Map of the X-ray sources compared in this paper. GC is the \object{Galactic Center} and the \object{Sun} is assumed to be 8.5 kpc far away from it.}
          \label{fig:map_sources}
    \end{figure}

\subsubsection{ISM abundances}
\label{section:discuss_abundances}

The abundances are displayed in Table \ref{table:ISM_abundances}, they are reported relative to the recommended proto-Solar values of \cite{Lodders}. We also show the abundances obtained by \cite{JuettII} towards the LMXB 4U 1820$-$303, which like GS 1826$-$238 is close to the Galactic center. Even more interesting is the comparison with the abundances estimated by \cite{kaastra} in the direction of the Crab nebula and the ones estimated by \cite{yao09} towards Cyg X$-$2, which are two different lines of sight in our Galaxy (see the map in Fig. \ref{fig:map_sources}).

The derived abundances slightly depend on the model used: the \textsl{pure-gas} model and the complete \textsl{gas+dust} model yield different results on Ne, Fe and Mg (Table \ref{table:rgs_fit}). First, their abundances are more uncertain than the oxygen abundance. Secondly, Fe and Mg are the most depleted elements: as dust grains give rise to absorption features different from gas particles, modeling edges of highly depleted elements provides different results by accounting or not for dust effects. Such deviations are probably strengthened since we observe through a high-density region: here high metal depletion is expected \citep{Williams96}.

All the heavy elements are over-abundant with respect to the proto-Solar values. Neon is over-abundant by a factor $\sim$ 1.7, as found by \cite{kaastra} in the \textit{XMM-Newton} observations of the Crab nebula. The Solar abundance of neon is probably under-estimated \citep[see][]{Lodders2009}, so our estimate may be not really different from the real Solar value. As we will show in the next paragraph, the metallicity gradient could also be responsible for part of the Ne over-abundance.

The reason for the over-abundances of O, Fe and Mg is rather different from the neon excess. First of all, we are able to measure both gas and dust contributions for O, Fe and Mg, without the risk of missing important fractions. Secondly, deviations in the abundances with respect to the average Galactic values are also due to their metallicity gradient. If $A(X)$ is the abundance of a certain element $X$ {in the vicinity of GS 1826$-$238,} we can write \citep[see][]{Esteban05},
\begin{equation}
\frac{A(X)}{A(X)_{\odot}} = 10^{\alpha_X \, (D_{GS} - \,D_{\odot})}
\label{eq:abundances}
\end{equation}
where $D_{\odot}$ and $D_{GS}$ are the Galactocentric radii of the Sun and GS 1826$-$238, {respectively $\sim$ 8.5 kpc and  $\sim$ 2 kpc}; $\alpha_X$ is the abundance gradient of the element {along the line of sight} and $A(X)_{\odot}$ is its abundance near the Sun. In this way, we can compare our estimates of abundance changes with the values predicted by the gradient estimates. Unfortunately, the gradient estimates in the literature are quite uncertain and are available only in a limited range of radii, i.e. between 4-16 kpc \citep[see][]{GradPedicelli}. Thus we can trust only in the abundance changes in $4-5$ kpc along our line of sight. Moreover, the column density estimated for each element refers to its integral along the line of sight, where we also expect a density increase towards the Galactic Center. {Because the density increases towards the Galactic center, the predicted abundance at the Galactocentric distance of GS 1826$-$238 should be close to the weighted average abundance along the sightline}:
\begin{itemize}
 \item Oxygen is over-abundant by about $20-30$ \% (see Table \ref{table:ISM_abundances}). According to \citet{Esteban05}, the oxygen gradient is $\alpha_{\rm O} = (-0.04 \pm 0.01) \,\rm{kpc}^{-1}$, which should provide an increment ({equation} \ref{eq:abundances}) of at least $\sim$ 32\% in the oxygen abundance. This is consistent with our estimate.
 \item The iron abundance {$\sim$ 1.20$-$1.54} (see Table \ref{table:ISM_abundances}) is near agreement with the {$\gtrsim$ 50\% increment} derived by the accepted iron gradient in the Galactic disk $\alpha_{\rm Fe} = (-0.06 \pm 0.02)\,{\rm kpc}^{-1}$ \citep{GradFriel, GradPedicelli}.
 \item Neon is over-abundant by more than 70\%. It is difficult to compare it with the Galactic gradient because this is quite uncertain, on average $\alpha_{\rm Ne} =  (-0.06 \pm 0.04)\,{\rm kpc}^{-1}$ \citep{Simpson1995, Maciel1999}. From {equation} (\ref{eq:abundances}) we predict a lower limit A(Ne) $\sim 1.3$. The sum of such a value to the revisited proto-Solar abundance \cite[$\Delta\,{\rm A(Ne)}\sim30$ \%, see][]{Lodders2009} provides A(Ne) $\sim 1.6$, which is fully consistent with our estimate. This result suggests that the neon over-abundance that we constrain is due to both the Galactic gradient and the previous under-estimate found in the literature.
 \item Nitrogen shows a steeper Galactic gradient of about $\sim -0.08\,{\rm kpc}^{-1}$ \citep{Gummersbach1998}, which provides $\Delta\,{\rm A(N)}\sim100$ \% ({equation} \ref{eq:abundances}) and agrees with our estimate (see Table \ref{table:ISM_abundances}). Of course, both of them are quite uncertain and we are not able to provide more information.
 \item Also magnesium should be at least twice the proto-Solar value as we found (see Table \ref{table:ISM_abundances}). However, it is difficult to compare our result with the Mg Galactic gradient found in the literature, because the results differ a lot in the literature \citep{Rolleston2000}. According to \citet{Gummersbach1998} $\alpha_{\rm Mg} \sim -0.08\,{\rm kpc}^{-1}$, which together with {equation} (\ref{eq:abundances}) implies abundance increment of at least 100\%. This is fully consistent with what we found.
\end{itemize}

{The iron and oxygen abundances towards 4U 1820$-$303 (Table \ref{table:ISM_abundances}) appear to disagree with our results despite its similar location near the Galactic center. \citet{JuettII} attribute this low iron abundance to depletion into dust grains in the interstellar medium and they also report that oxygen could be middle-depleted. In Sect. \ref{sec:dust_analysis} we have shown that iron is among the most depleted elements in the ISM, oxygen is also partially depleted and a pure-gas model cannot reproduce all the ISM spectral features. This indicates that the differences between the Fe and O abundances are due to the capability of our gas+dust model to measure the contribution from solid phases, which are absent in the pure-gas model of \citet{JuettI,JuettII}.}

{The \object{Crab} nebula is relatively close to the Solar system, i.e. $\sim$ 2 kpc, but opposite to the Galactic center (see Fig. \ref{fig:map_sources}). Thus we expect abundances close to the proto-Solar values of \citet{Lodders}, which is just what \citet{kaastra} found.}

{LMXB Cyg X-2 is also far away from the center of the Galaxy and about 10 kpc away from the Sun. The abundances estimated for this source are lower than those measured towards the Crab and GS 1826$-$238 (see Table \ref{table:ISM_abundances}). This agrees with the assumed abundance gradients we have discussed.}

In summary, the differences with respect the proto-Solar abundances that we estimate are consistent with the literature. The increase of metallicity towards the center of the \object{Galaxy} should be due to evolutionary effects like supernovae explosions, which enrich the ISM with heavy elements, especially in the higher density region of the bulge and the disk of our Galaxy. A deeper analysis is required: we need to study more sources, even in the same region, and further improve our models in order to account for every contribution to the column densities.

\section{Conclusion}
\label{conclusion}

We have presented a complete treatment of the interstellar medium towards the low-mass X-ray binary GS 1826$-$238, which is a bright X-ray source near the Galactic center. We have shown that in the line of sight the ISM is composed of a complex mixture of a multi-phase gas, dust and molecules. 

The gas is almost neutral and the ionization degree is about 5\%. Significant fractions of the column density of some heavy elements are in the form of molecules or dust grains: at least 10\% of oxygen, 50\% of magnesium and 90\% of iron. Such a solid phase should consist of a mixture of silicates ($\gtrsim$ 60\%), CO ($\lesssim$ 10 \%), ice ($\lesssim$ 20 \%) and other iron oxides ($\lesssim$ 10 \%).

We have found over-abundances for all the elements for which we have been able to measure the individual column density. The Ne over-abundance that we estimate is consistent with other recent work in different lines of sight, such as towards the Crab \citep{kaastra}, suggesting that the Solar value is underestimated. However, our agreement with the predicted Ne gradient in the Galaxy is also consistent with an abundance increase due to stellar evolution: towards the Galactic center there are high-density regions with evolved star that could have enriched the ISM with heavy elements such as neon.

Differently from the previous X-ray spectroscopy work, we have found over-abundances for oxygen ($1.2$), iron ($1.4$) and magnesium ($2.4$). These elements are also in the form of dust and molecules. Thus our estimates are partially due to the fact that we are able to measure also the contributions from the solid phase. The abundance of metals is in agreement with the metallicity gradients and shows the chemical inhomogeneity of the interstellar medium.

The diagnostic of the ISM constituents fits the predicted models for both its thermal and chemical structures, showing a good agreement with the current state of art \citep{ferriere}. The dust column is consistent with the multi-wavelength, X-ray versus IR, observations. All this supports our research method and justifies new observations and analysis of other sources in different lines of sight. Such analysis, indeed, can provide a better mapping of the ISM and a deeper study of its chemical composition, together with its role in the evolution of the entire Galaxy.

\begin{acknowledgements}
We are grateful to Jean in't Zand for the useful discussion and clarification about the physics of low-mass X-ray binaries. This work is based on observations obtained with \textit{XMM-Newton}, an ESA science mission with instruments and contributions directly funded by ESA Member States and the USA (NASA). SRON is supported financially by NWO, the Netherlands Organization for Scientific Research.
\end{acknowledgements}

\bibliographystyle{aa}
\bibliography{bibliografia} 



\begin{appendices}

\section{{\sl Amol}: oxygen edge molecules absorption model}
\label{appendix:amol}


The {\sl amol} model calculates the transmission of various molecules. 
Presently only the oxygen edge is taken account of. Updates of this model,
once made, will be reported in the manual of SPEX. The following compounds are presently taken into account (see
Table~\ref{tab:amol}).

\begin{table}[!h]
\caption{Molecules present in the {\sl amol} model.}
\label{tab:amol}
\smallskip
\centerline{
\begin{tabular}{rllc}
\hline
nr & name             & chemical formula                    & reference \\
\hline
 1 & molecular oxygen & O$_2$                               & a \\
 2 & carbon monoxide  & CO                                  & a \\
 3 & carbon dioxide   & CO$_2$                              & a \\
 4 & laughing gas     & N$_2$O                              & a, b \\
 5 & water            & H$_2$O                              & c \\
 6 & crystalline ice  & H$_2$O                              & d \\
 7 & amorphous ice    & H$_2$O                              & d \\
 8 & cupric oxide     & CuO                                 & e \\
 9 & nickel monoxide  & NiO                                 & e \\
10 & iron oxide       & Fe$_{1-x}$O                         & e \\
11 & magnetite        & Fe$_3$O$_4$                         & e \\
12 & hematite         & Fe$_2$O$_3$                         & e \\
13 & eskolaite        & Cr$_2$O$_3$                         & e \\
14 & andradite        & Ca$_3$Fe$_2$Si$_3$O$_{12}$          & e \\
15 & acmite           & NaFeSi$_2$O$_6$                     & e \\
16 & franklinite      & Zn$_{0.6}$Mn$_{0.8}$Fe$_{1.6}$O$_4$ & e \\
17 & chromite         & FeCr$_2$O$_4$                       & e \\
18 & ilmenite         & FeTiO$_3$                           & e \\
19 & perovskite       & CaTiO$_3$                           & e \\
20 & olivine          & Mg$_{1.6}$Fe$_{0.4}$SiO$_4$         & e \\
21 & almandine        & Fe$_3$Al$_2$(SiO$_4$)$_3$           & e \\
22 & hedenbergite     & CaFeSi$_2$O$_6$                     & e \\
23 & hercynite        & FeAl$_2$O$_4$                       & e \\
\hline\noalign{\smallskip}
\end{tabular}
}
references: \\
$^a$ \cite{barrus79}, 0.5-0.75 eV resolution\\
$^b$ \cite{wright74}, 0.5 eV resolution \\
$^c$ \cite{hiraya01}, 0.055 eV resolution \\
$^d$ \cite{parent02}, 0.1 eV resolution \\
$^e$ \cite{vanaken98}, 0.8 eV resolution \\
\end{table}

The chemical composition of these minerals was mainly taken from the Mineralogy
Database of David Barthelmy\footnote{http://webmineral.com/}. We take the cross-sections from the references as listed in Table~\ref{tab:amol} in the energy interval where these are given, and use the cross section for free atoms \cite{verner95} outside this range. \cite{vanaken98} do not list the precise composition of iron oxide. We assume
here that $x=0.5$.

Some remarks about the data from \cite{barrus79}: not all lines are given in their tables, because they suffered from instrumental effects (finite thickness absorber combined with finite spectral resolution). However, Barrus et al. have estimated the peak intensities of the lines based on measurements with different column densities, and they also list the FWHM of these transitions. We have included these lines in the table of cross sections and joined smoothly with the tabulated values.

For N$_2$O, the fine structure lines are not well resolved by Barrus et al. Instead we take here the relative peaks from \cite{wright74}, that have a relative ratio of 1.00 : 0.23 : 0.38 : 0.15 for peaks 1, 2, 3, and 4, respectively. We adopted equal FWHMs of 1.2 eV for these lines, as measured typically for line 1 from the plot of \cite{wright74}. We scale the intensities to the peak listed by  \cite{barrus79}. Further, we subtract the C and N parts of the cross section as well as the oxygen 2s/2p part, using the cross sections of \cite{verner95}.  At low energy, a very small residual remains, that we corrected for by subtracting a constant fitted to the 510--520 eV range of the residuals. The remaining cross section at 600 eV is about 10 \% above the Verner cross section; it rapidly decreases; we approximate the high-E behavior by extrapolating linearly the average slope of the ratio between 580 and 600 eV to the point where it becomes 1.
\end{appendices}



\end{document}